\documentclass[apjl,twocolumn]{emulateapj_mod}
\usepackage{epsfig,apjfonts,mathptmx}

\def\gtsima{$\; \buildrel > \over \sim \;$}
\def\ltsima{$\; \buildrel < \over \sim \;$}
\def\prosima{$\; \buildrel \propto \over \sim \;$}
\def\gsim{\lower.5ex\hbox{\gtsima}}
\def\lsim{\lower.5ex\hbox{\ltsima}}
\def\simgt{\lower.5ex\hbox{\gtsima}}
\def\simlt{\lower.5ex\hbox{\ltsima}}
\def\simpr{\lower.5ex\hbox{\prosima}}

\def\h1{$h^{-1}$}
\def\eeq{\end{equation}}
\def\beq{\begin{equation}}

\submitted{Submitted 13 May 2005; Accepted 20 July 2005}

\shorttitle{Multi-$\lambda$ observations of $z=2$ galaxies in GOODS-N}
\shortauthors{E. Daddi et al.}

\journalinfo{Astrophysical Journal Letters in press}
\begin{document}

\title{The population of B$\lowercase{\rm z}$K selected ULIRG$\lowercase{\rm s}$ at 
$\lowercase{z}\sim2$}

\author{E. Daddi\altaffilmark{1,2},
	M. Dickinson\altaffilmark{1},
	R. Chary\altaffilmark{3},
	A. Pope\altaffilmark{4},
	G. Morrison\altaffilmark{1},
	D.M. Alexander\altaffilmark{5},
	F.E. Bauer\altaffilmark{5},
	W.N. Brandt\altaffilmark{6},
	M. Giavalisco\altaffilmark{7},
	H. Ferguson\altaffilmark{7},
	K.-S. Lee\altaffilmark{7},
	B.D. Lehmer\altaffilmark{6},
	C. Papovich\altaffilmark{8},
	A. Renzini\altaffilmark{9}
}

\altaffiltext{1}{National Optical Astronomy Observatory,
950 N. Cherry Ave., Tucson, AZ, 85719} 
\altaffiltext{2}{{\em Spitzer} Fellow; edaddi@noao.edu}
\altaffiltext{3}{{\em Spitzer} Science Center, Caltech, MS 220-6, CA 91125}
\altaffiltext{4}{Department of Physics \& Astronomy, University of British Columbia, Vancouver, BC, V6T 1Z1, Canada}
\altaffiltext{5}{Institute of Astronomy, Madingley Road, Cambridge CB3 0HA, UK}
\altaffiltext{6}{Department of Astronomy and Astrophysics, 525 Davey Laboratory, Pennsylvania State University, University Park, PA 16802}
\altaffiltext{7}{Space Telescope Science Institute, 3700 San Martin Drive, Baltimore, MD 21218}
\altaffiltext{8}{Steward Observatory, University of Arizona, Tucson, AZ 85721}
\altaffiltext{9}{ESO, Karl-Schwarzschild-Strasse 2, Garching 85748, Germany}

\begin{abstract} We investigate the multi-wavelength emission of
$BzK$ selected star forming galaxies at $z\sim2$ in the Great Observatories 
Origins Deep Survey (GOODS) North region. 
Most (82\%) of the sources are individually detected at 
24$\mu$m in the Spitzer MIPS imaging, and one fourth (26\%) in the VLA radio 
data.
Significant detections of the individually undetected objects are obtained 
through stacking in the radio, submm and X-ray
domains. The typical
star forming galaxy with stellar mass $\sim10^{11}M_\odot$ at $z=2$ is an Ultra-luminous Infrared Galaxy (ULIRG),
with $L_{\rm IR}\sim1$--2$\times10^{12}L_\odot$ and star formation rate
$SFR\approx200$--300$M_\odot$yr$^{-1}$, implying a comoving
density of ULIRGs at $z=2$
at least 3 orders of magnitude above the local one. $SFR$s derived from the 
reddening corrected UV luminosities agree well, on average, with the longer
wavelength estimates. The high 24$\mu$m detection rate suggests a relatively
large duty cycle for the $BzK$ star forming phase, 
consistently with the available independent measurements of the space
density of passively evolving galaxies at $z>1.4$. 
If the IMF at $z=2$ is similar to the local one, and in particular is not
a top-heavy IMF, 
this suggests that a substantial fraction of  the high mass tail 
($\simgt10^{11}M_\odot$) of the galaxy stellar mass function was completed by $z\approx1.4$.
\end{abstract}

\section{Introduction}

The rate at which stars in massive galaxies were 
assembled is a crucial measurement for
the characterization of  galaxy formation.
Hierarchical clustering models
have traditionally favored galaxy formation at 
quiescent rates (e.g. Cole et al. 2001). 
Formation of stars at high
rates in massive galaxies would qualitatively match monolithic formation 
scenarios (Eggen et al. 1962). 
Intense star formation at high redshift has been established for submm
selected galaxies. These are relatively extreme
objects, 
with space densities nearly two orders of magnitude smaller than local 
massive galaxies (e.g., Scott et al. 2002). 
Extending the census to less extreme star formation rate ($SFR$)
levels is necessary, but
obtaining reliable $SFR$ estimates for high-redshift galaxies is a 
challenging endeavor. The widespread presence of dust, absorbing 
the light of high mass stars and re-radiating it at longer wavelengths,
complicates the immediate use of the UV luminosity as a $SFR$ indicator, as
uncertain large 
corrections are required. A multi-wavelength approach, although 
observationally demanding, can yield more robust measures of $SFR$.

The major growth and assembly of galaxies' stellar mass is observed
during the epoch between redshifts $1<z<3$ (Dickinson et 
al. 2003; Rudnick et al. 2003). 
Daddi et al. (2004b), on the basis of the highly
complete spectroscopic database of the K20 survey (Mignoli et al. 2005), 
showed that
a relatively clean, efficient and complete (reddening independent)
selection of massive
galaxies in the above redshift range can be obtained  by selecting 
outliers in a $(B-z)$ vs. $(z-K)$ diagram. For $z\sim2$ objects
with $K<20$, selected in this way, Daddi et al. (2004a;b; hereafter
D04a and D04b) 
derived high masses and $SFR$s, suggesting they 
are the progenitors of local massive spheroids caught during their
phase of major assembling. The $SFR$s derived 
from the dust-corrected UV luminosities are however uncertain. 

In this letter, we have taken advantage of the $BzK$ selection to assemble a 
statistical sample of $K<20$ (Vega) massive star forming galaxy candidates
at $1.4<z<2.5$ in the  Great Observatories
Origins Deep Survey (GOODS) North field.  
Deep 24$\mu$m observations, 
obtained with MIPS on board of the Spitzer Space 
Telescope (SST), were used to study their rest-frame mid-IR emission properties.
Deep X-ray, submm and radio data allow us to build a panchromatic view
of their spectral energy distributions (SEDs), 
and to shed light on their nature and 
star formation activity. We discuss the implications for the assembly process
of massive galaxies.
We use a Salpeter initial mass function (IMF) from 0.1 to 100 $M_\odot$, and a WMAP cosmology with
$\Omega_\Lambda, \Omega_M = 0.73, 0.27$, and
$h = H_0$[km s$^{-1}$ Mpc$^{-1}$]$/100=0.71$.

\section{B$\lowercase{z}$K selection of $z\sim2$
galaxies in GOODS-North}

The GOODS-North field has been observed in the $K$-band
with the Flamingos camera at the Mayall 4-m NOAO telescope.
About 4--7 hours integration were collected with 
average 1.2$''$ seeing, over each of two
contiguous pointings, reaching $5\sigma$ limits of $K\sim20.5$ (Vega)
for point sources. Calibration was done using objects in common with 2MASS.
Observations in the $B$- and $z$- bands were obtained at the Subaru telescope 
with 
Suprime-Cam with $<1''$ seeing, and are described in Capak et al. (2004).
Sources were selected over an area of 154 arcmin$^2$ to $K<20$ (Vega), 
thus matching the limit 
where the  $BzK$ criterion is currently calibrated by the K20 survey (D04b). 
The Capak et al. (2004) zeropoints were slightly adjusted, 
and a small color term was applied to the $B$-band  magnitudes
because the Subaru $B$-band filter is
redder than the one used in D04b.
169 $z\sim2$ star forming galaxy candidates were selected having
$BzK\equiv (z-K)_{AB}-(B-z)_{AB} >-0.2$. We are explicitely
excluding from the analysis 
the 13 objects with $BzK<-0.2$ and $(z-K)_{AB}>2.5$ that are
candidate passively evolving galaxies at $z>1.4$.
However, the shallower depth
of the $B$-band data (compared to those in D04b)
prevents a clean separation of $z>1.4$ passive and star forming
galaxies among sources with the reddest $z-K$ colors.
We might expect up to 10-15 genuine
passively evolving $z>1.4$ galaxies among the $BzK>-0.2$ objects,
counting the sources with no or low-significance B-band detection.
38/169 (22\%) objects 
were discarded as likely AGN-dominated, because
detected in the hard X-ray band (Alexander et al. 2003). The
AGN contamination that we recover here is
higher than in D04b, owing to
the deeper X-ray data. 
The surface density of the 131 non X-ray bright
$BzK>-0.2$ objects in GOODS-N is $0.85\pm0.07$ arcmin$^{-2}$
(Poisson), consistent with that measured in the K20 survey.  On the basis of
the K20 survey results (D04b),
we assume in the
following that the 131 galaxies with $BzK>-0.2$ are mostly at
$1.4<z<2.5$, with a relatively flat redshift distribution in that
interval and with a $\simlt10$\% contamination of objects at
$1<z<1.4$, and that our sample is complete for star forming galaxies
with $K<20$ at $1.4<z<2.5$. This is also
supported by our photometric redshifts, and by the 
limited amount of spectroscopic redshift currently available for $K<20$, $BzK$
sources in GOODS-N.

\section{Multi-$\lambda$ observations of B$\lowercase{z}$K selected galaxies}

{\em Rest frame UV}: 
The calibrations of D04b were used for estimating stellar masses, 
reddening and $SFR$ of $BzK$ selected galaxies from their observed properties 
in the optical/IR. The average 
mass of our $z=2$ object is $1.0\times10^{11}M_\odot$ and the sample is 
complete above $\sim10^{11}M_\odot$ for  $1.4<z<2.5$. 
The median $B-z=1.50$ color
translates to a reddening of $E(B-V)=0.40$ for the Calzetti et al. (2000)
law. This median reddening  is similar
to our estimate of the $E(B-V)$ upper limit for the  UV criteria
of Steidel et al. (2004) for selecting $z\sim2$
galaxies, and we estimate that roughly 50\% of the $z\sim2$ $BzK$
star forming
galaxies with $K<20$ would be missed when selecting in the UV. We find
an average $SFR\sim220\ M_\odot$yr$^{-1}$ for our sample, 
based on the estimate of the reddening corrected 1500\AA\ rest frame luminosity,
with a median (average) correction factor of 46 (118). The ratio of the averages of
the corrected and uncorrected $SFR$s is 42.

{\em MIPS 24$\mu$m}:
Deep 24$\mu$m observations of GOODS-North were obtained with SST MIPS,
as a part of the GOODS
Legacy Program (M. Dickinson et al., in preparation),
for a total of 10.4 hours exposure time per sky pixel.
Sources were detected in the 24$\mu$m data using SST IRAC data as prior 
positions, in order to improve source deblending (R. Chary et al., 
in preparation).  We cross-correlated
the 131 $BzK$ selected galaxies to MIPS objects using a 2$''$ search 
radius (expected false matching rate of order of a few \%). 107/131
(82\%) of the $z\sim2$ galaxies are matched to a
MIPS counterpart. The 24$\mu$m flux 
($f_{24}$) ranges from 300-500 $\mu$Jy for the brightest sources to the
typical $3\sigma$ limits in the range of 15--30$\mu$Jy (depending on
position) for undetected sources. 
The median $f_{24}$ is 110$\mu$Jy,
including non detected sources. The average
is 127--124$\mu$Jy, depending on whether for the undetected sources
we use the $3\sigma$ upper limit or assign zero flux to them.
A Kendall's $\tau$ test
indicates that $f_{24}$ correlates with the $K$-band flux, with
a  99\% level of significance. 
Comparing to the full GOODS-N MIPS 24$\mu$m catalog,
$BzK$ star forming galaxies account
for about 7--10\% of 24$\mu$m selected sources above 50 to 300
$\mu$Jy, with the fraction varying slightly with 24$\mu$m
limiting flux.  These figures place lower limits to the
fraction of $z>1.4$ galaxies in 24$\mu$m selected galaxy samples,
which are consistent with the predictions of Chary \& Elbaz (2001; 
hereafter CE01), i.e. 16\%. 
At $1.4<z<2.5$ the 24$\mu$m observations probe rest-frame wavelengths
of 7--10$\mu$m, the mid-IR region dominated by strong PAH emission
features and silicate absorption.  
In order to constrain the level of IR 
luminosity ($L_{\rm IR}\equiv L_{8-1000\mu{\rm m}}$), hence $SFR$, required to
reproduce the observed $f_{24}$ levels, 
we used the models from CE01, which
provide $L_{\rm IR}$-dependent templates calibrated from the local SEDs of 
IR luminous galaxies. For a given $L_{\rm IR}$, 
these models predict a factor of $\approx 5$
decrease of $f_{24}$ from $z=1.4$ to $z=2.5$. 
Assuming a flat
redshift distribution within $1.4<z<2.5$ for the $BzK>-0.2$ selected galaxies, 
models with
$L_{\rm IR}=1.7\times10^{12}L_\odot$ are required to reproduce the 125$\mu$Jy
average flux level\footnote[1]{
Note that we are studying the 24$\mu$m properties of a near-IR selected
sample of galaxies, not a 24$\mu$m selected sample. Its redshift
distribution is expected to be flat within $1.4<z<2.5$, independent of
the behavior of the 24$\mu$m flux with redshift. As we are detecting
82\% of the objects, the typical flux that we detect corresponds to the
source at typical redshift.}. This corresponds to an average 
$SFR\sim300\ M_\odot$yr$^{-1}$ (Kennicutt et al. 1998).
Using instead the empirical SED of Arp~220,
calibrated by ISO observations of the PAH features region,
one would expect a relative $f_{24}$ peak around $z\sim2$ and minima
toward $z=1.4$ and $z=2.5$. This is due to strong 9.7$\mu$m
silicate absorption, a feature which has been observed to be quite
common at $z\sim2$ in sources with $L_{\rm IR}\simgt10^{13}L_\odot$
(e.g., Yan et al. 2005). 
The $f_{24}$ expected from Arp~220 when averaged within
$1.4<z<2.5$ is about 30$\mu$Jy.  For the Arp~220 IR
luminosity of $1.5\times10^{12}L_\odot$, this in turn implies
$L_{\rm IR}\sim6\times10^{12}L_\odot$ for the typical $BzK$ star forming 
galaxy at
$z=2$, or very large $SFR\sim1000\ M_\odot$yr$^{-1}$ (this high value is
disfavored by the other measurements, see below).

{\em Radio 20cm}: 34/131 (26\%) of the $BzK>-0.2$ galaxies are individually 
detected
with $S/N>3$ in deep VLA 
radio data (Richards 2000; reprocessed by G. Morrison et al., in preparation), 
with fluxes at 1.4~GHz in the range of 20--160~$\mu$Jy (when excluding a
radio galaxy with 1mJy flux), and an average of 37~$\mu$Jy. 
Stacking of the 
94 undetected sources\footnote[2]{
To stack the radio data, sub images were
extracted at the locations of the non-detected sources.
These were corrected for
VLA primary beam attenuation and combined using a weighted
average. The integrated flux density and error were computed with 
the AIPS task
JMFIT, which modeled the stacked emission using an elliptical Gaussian.}, 
after discarding 3 objects close to unrelated radio 
sources, yields a $S/N=8$ detection for an estimated
average flux density of $10\pm2$~$\mu$Jy.
When adding to this the individual detections we derive an average flux 
density of about 17$\mu$Jy for the full sample of 131 objects. 
Using a radio spectral index $\alpha=-0.8$ ($f_{\nu}\propto\nu^{-\alpha}$) we derive a luminosity 
of $3.6\times10^{23}$ W~Hz$^{-1}$ for an average $<z>=1.9$, which implies
$L_{\rm IR}\sim1.2\times10^{12}L_\odot$ and
$SFR\sim210\ M_\odot$yr$^{-1}$ (Yun et al. 2001; Kennicutt et al. 1998).

\begin{figure}[ht]
\centering 
\includegraphics[width=8.2cm]{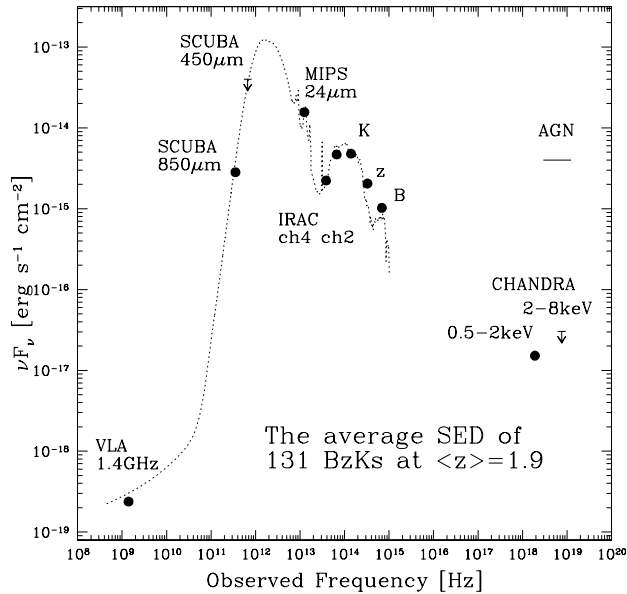}
\caption{The average multi-wavelength emission of 131 $BzK$ 
selected star forming
galaxies at $1.4<z<2.5$. The model shown here (from CE01)
has $L_{\rm IR}=1.7\times10^{12}L_\odot$ and is redshifted to $z=1.9$. Measurements
in SST IRAC ch2 (4.5$\mu$m, $AB=20.78$ mag) and ch4 (8.0$\mu$m, $AB=20.99$ mag)
were obtained from the data
of M. Dickinson et al., in preparation. The horizontal bar on the right shows
the 2-8~keV emission that would be expected from a typical AGN having the
mid-IR flux observed for our objects.
}
\label{fig:multi}
\end{figure}

{\em SCUBA 850$\mu$m}: Only one of the 131 sources appears in the 
list of Pope et al. (2005) with a flux of 4mJy. In order to estimate
the average 850$\mu$m flux of the undetected objects we used the 
Borys et al. (2003) HDF-N SCUBA supermap. Some 97 MIPS-detected 
objects were stacked\footnote[3]{We are 
using only MIPS-detected objects to maximize the signal to noise  by avoiding
passive galaxy contaminants, 
while the possible star forming galaxy contaminants at $1<z<1.4$ would not alter
significantly the stacking because of the mild dependence between 850$\mu$m
flux with redshift.}, after
excluding those separated by less than a 
SCUBA beam halfwidth (7") from any known SCUBA
detection.
As the depth of the SCUBA data vary 
considerably over GOODS-N, we used
weights in the stacking based on the local noise.
The average 850$\mu$m signal is $1.0\pm0.2$mJy. We checked
that this 5-sigma detection is robust using  Monte-Carlo simulations.
Scaling down the result by 18\% to account for the MIPS undetected
objects, the average 850$\mu$m flux corresponds to 
$L_{\rm IR}\sim1.0\times10^{12}L_\odot$, both for the CE01 templates and when using
Arp~220, or to $SFR\sim170\ M_\odot$yr$^{-1}$. At 450$\mu$m we obtain a 3$\sigma$
upper limit of 6mJy, consistent with the models.

{\em Chandra soft and hard X-rays}: 
Is the detected IR emission AGN or starburst powered~? 
At $z\sim2$ the Chandra hard 2--8~keV
band is sensitive to highly penetrating rest-frame 
X-rays up to about 20~keV, which can escape large column densities
up to about $10^{24}$~cm$^{-2}$.
The lack of hard
X-ray detection in the Chandra 2 Ms exposure should ensure that
bright AGNs are generally not included in our sample. 
The typical upper limits to the 2--8~keV
flux of order $<3\times10^{-16}$ erg s$^{-1}$ cm$^{-2}$ coupled to the
typical 24$\mu$m fluxes measured, imply lower limits to the 24$\mu$m
to 2--8~keV flux ratios, for single objects, about one full order of
magnitude larger than expected from AGNs (see e.g. Rigby et al
2004). This suggests that the 24$\mu$m emission in our sample is
powered by star formation in most cases. 
While we have discarded all direct hard
X-ray detections from our sample, 6/131 (4.5\%) of the objects are detected
in the soft-band only, with fluxes $\simlt10^{-16}$~erg~s$^{-1}$~cm$^{-2}$
(Alexander et al. 2003).
Four of these are detected also in the radio, and one is the only SCUBA
detection. These soft sources are most likely
the most extreme starbursts in the $BzK>-0.2$ sample, similar to object ID\#5
described in D04a. 
We stacked the X-ray undetected sources, considering only
the GOODS-N region within
6 arcmin of the 2 Ms Chandra data aim point where the sensitivity is the
highest, and avoiding all objects closer than 3 Chandra PSF from known X-ray
sources. The stacked images result in a 8.5$\sigma$ detection 
in the soft 0.5-2~keV band, for a flux of 
$1.0\times10^{-17}$~erg~s$^{-1}$~cm$^{-2}$.
The non detection in the hard 2-8~keV band constrains the photon spectral index
$\Gamma>1.0$ at the 3-sigma level, and implies an average 24$\mu$m to 2-8~keV
flux ratio over two orders of magnitude above what is typical for AGN at $z\sim2$
(Rigby et al. 2004). If AGN are present they would have to be heavily Compton
thick.
Using $\Gamma=2.0$ as appropriate for 
starbursts implies a rest-frame luminosity in the 
2--10~keV range of $3.4\times10^{41}$~erg~s$^{-1}$ for $<z>=1.9$.
Adding back the individual soft X-ray detections would increase the 2--10~keV
luminosity to about $5\times10^{41}$~erg~s$^{-1}$. 
This is a factor of two lower than estimated by D04b
for K20 $BzK$ sources.
Using the Ranalli et al. (2003) calibration this translates into an average 
$SFR\sim100\ M_\odot$yr$^{-1}$, or $SFR\sim500\ M_\odot$yr$^{-1}$ 
if using the Persic
et al. (2004) calibration. The large difference is due to the relative 
expected importance of high versus low mass X-ray binaries.
The latter calibration appears more appropriate 
when dealing with IR luminous sources with $L_{\rm IR}\simgt10^{12}L_\odot$
(Persic et al. 2004). 

\section{Discussion}

The multi-wavelength SED (Fig.~\ref{fig:multi}) of $BzK$ selected $z=2$ 
star forming galaxies consistently indicates an
average $L_{\rm IR}\sim1$--2$\times10^{12}L_\odot$ and 
$SFR\sim200$--300$\ M_\odot$yr$^{-1}$,
supporting the earlier claims of D04a;b and 
implying that the local correlations (e.g., the mid and far-IR to radio
correlations) hold, for the average $BzK$ galaxy,
also at $z\approx2$. While the X-ray based estimate appears
the least accurate, due mainly to the large uncertainties in its calibration, 
the X-ray properties clearly support that the mid to far-IR emission of 
these sources is indeed dominated by vigorous star formation and not by nuclear activity.
The inferred average $L_{\rm IR}\simgt10^{12}L_\odot$ implies
that the typical $BzK$ selected star forming galaxy is
an Ultra Luminous IR Galaxy (ULIRG). 
Morphology from HST ACS indeed suggests that, in many cases, these
$z=2$ galaxies are assembling through merging (D04a),
similarly to local ULIRGs.
Using the volume at $1.4<z<2.5$ ($5.7\times10^5$~Mpc$^{3}$), we can put a lower limit to the spatial density of ULIRGs at $z\sim2$
of about 1--2$\times10^{-4}$ Mpc$^{-3}$. This is about 3 orders of
magnitude higher than at $z\sim0.1$ (Sanders et al. 2003), and a factor of 2--3
higher than at $z=1$ (Le Floc'h et al. 2005).
While ULIRGs are exceptional objects for the local
universe, they appear to be the norm among massive $z\approx2$ star forming
galaxies.

\begin{figure}[ht]
\centering 
\includegraphics[width=8.2cm]{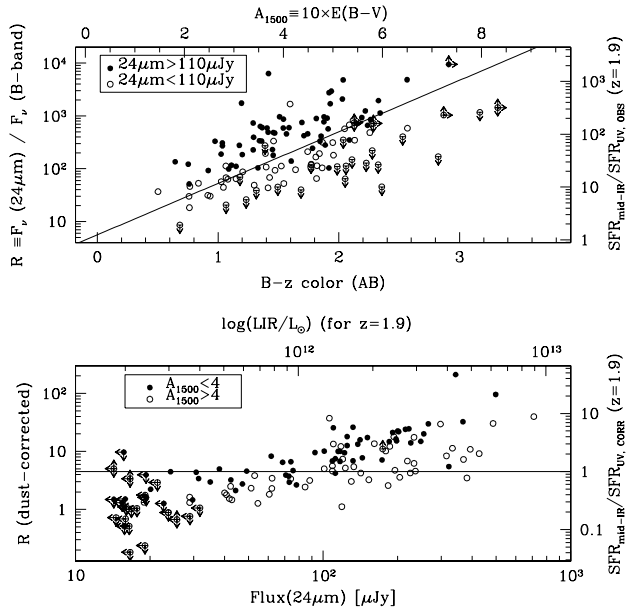}
\caption{({\em Top panel}). The
ratio of 24$\mu$m to B-band flux for $K<20$ $BzK$ selected $z=2$
star forming galaxies in GOODS-North
as a function of $B-z$ color  (i.e. dust extinction at 1500\AA, as inferred
using a Calzetti et al. 2000 law). 
The diagonal line shows the expected ratio for the average $z=1.9$ in the case
that the UV and mid-IR trace the same amount of star formation. Solid and empty symbols are for sources with $f_{24}$ above or below the median, respectively.
({\em Bottom panel}). The dust extinction corrected 24$\mu$m to B-band flux
ratio, as a function of the 24$\mu$m flux. Solid and empty symbols are for sources with UV reddening below or above the median, respectively.
In both panels, values on axis labeled with $L_{\rm IR}$ or $SFR$ are for an 
average $z=1.9$ and were computed using the CE01 model reproducing the average SED
(Fig~\ref{fig:multi}).
}
\label{fig:B24pol}
\end{figure}

From the X-ray we also
derive a fairly low X-ray to optical flux ratio of 
${\rm log(f_{0.5-2~keV}/f_R)\sim-1.7}$, which suggests that also
the rest frame UV emission of these $z=2$ galaxies
is dominated by the emission of stars. The average $SFR$ inferred from
the dust-corrected UV luminosity agrees well with the
longer wavelength estimates.
Fig.~\ref{fig:B24pol} (top) shows the ratio of 24$\mu$m to $B$-band flux
densities as a function of the rest-frame UV galaxy colors. 
Apart from K-correction effects, this observed
mid-IR to UV luminosity ratio measures the ratio of
dust-extinguished to relatively unobscured $SFR$.  The Kendall's $\tau$
test detects a correlation at $>99.9$ confidence level. A similar trend 
is found when using the radio flux instead of $f_{24}$.
This confirms that, in general, the red UV continua of
$BzK$ selected $z=2$ star forming galaxies are indeed due 
to dust reddening. The diagonal
line plotted in Fig.~\ref{fig:B24pol} (top)
shows the expected scaling in the case that the UV and mid-IR trace the same 
amount of $SFR$.
The average ratio and its dependence
with the UV color are in line with what is expected. 
Sources with higher 24$\mu$m flux, however,
tend to have systematically larger ratios, and vice-versa.
This can be more clearly seen in Fig.~\ref{fig:B24pol} (bottom), where we
plot the 24$\mu$m to $B$-band flux ratio (with the $B$-band flux
here being corrected for dust reddening) as a function of the 24$\mu$m flux:
a clear trend with 24$\mu$m flux is present. 
The expected K-correction term is small, and this plot 
suggests that sources
with higher $L_{\rm IR}$ have progressively
larger fractions of the $SFR$ that cannot be 
recovered from the UV luminosity even after
correcting for dust reddening. However,
such a strong trend is not reproduced using the radio data for the radio detected 
objects. Fig.~\ref{fig:B24pol} (bottom) also shows that, at fixed $f_{24}$, 
the
sources with bluer UV continua tend to have higher
corrected 24$\mu$m to $B$-band
flux ratio by a factor of 2 on average. 

Vigorous starbursts are present within the $BzK$ selected star forming
galaxies at $z=2$, which have also fairly large typical stellar masses of
$\sim10^{11}$M$_\odot$ for $K<20$ (Vega). 
These masses would grow even larger 
as a result of star formation, depending on the duty cycle of the 
star formation event. 
The high detection rate (82\%)
at 24$\mu$m supports the possibility that high $SFR$s among these $K<20$
sources are sustained for a substantially long amount of time during $1.4<z<2.5$. 
When limiting to a common stellar mass threshold, e.g. $\simgt10^{11}M_\odot$, the space
densities of passively evolving galaxies is comparable to that of vigorous starbursts
within $1.4<z<2$, and perhaps much smaller at $2<z<2.5$ (Daddi 
et al. 2005; Kong et al. 2005; McCarthy et al. 2004). This would suggest that the average duty 
cycle is at least 50\%, and likely
more. 
The Universe ages from 2.6 Gyr to 4.6 Gyr between $z=2.5$ and 1.4, and about
1~Gyr is still available on average per galaxy before $z=1.4$, implying a
continuation of the star formation event for order 0.5~Gyr or more, 
on average, and that typically these objects have been active for a similar
or larger amount of time before observations.
This is also consistent with the typical age of the present SF event inferred from the 
optical/IR SED fitting with constant star formation rate models, which is mainly based 
on the strength of the observed Balmer break
(about 0.7~Gyr; D04a; see also Shapley et al. 2005). 
Therefore, by $z=1.4$ the typical mass of these 
galaxies will have roughly doubled, reaching $\sim2\times10^{11}$M$_\odot$, on average.
From their observed space densities ($\sim2.3\times10^{-4}$~Mpc$^{-3}$)
and $SFR$s, 
we infer that the integrated stellar mass density formed within $BzK>-0.2$
galaxies with $K<20$ in the 2~Gyr time within $1.4<z<2.5$ is 
$\sim 10^{8}M_\odot$ Mpc$^{-3}$, independent of the duty cycle. 
This is only $\sim20$\% of the local total stellar mass density,  but is
comparable to the local stellar 
mass density for objects with stellar mass $>2\times10^{11}$M$_\odot$
(Cole et al. 2001). 
If the IMF at $z=2$ is similar to the local one, and in particular is not
a top-heavy IMF, 
this suggests that by $z\approx1.4$ the assembly of the
high-mass tail ($>10^{11}$M$_\odot$) of present day's
galaxies mass function was, in a significant part, completed.           
This is also supported by
the measurements, quoted above, of a comparable space densities of 
old and passive 
galaxies with similar masses to the star-forming $BzK$s
 already existing at $1.4<z<2$, 
and with the evidences (Papovich et al. 2005) of strongly declining specific $SFR$s for $z<1.4$ massive galaxies.
%\vspace{0.1truecm}
\acknowledgments
We thank the many other members of the GOODS team
who have helped to make these observations possible.
We are grateful 
to Colin Borys for his work on the HDF-N SCUBA supermap and to the 
anonymous referee and to the editor, John Scalo, for useful comments.
ED gratefully acknowledges NASA support through the Spitzer 
Fellowship Program, award 1268429.
Support for this work, part of the Spitzer Space Telescope Legacy
Science Program, was provided by NASA through Contract Number 1224666
issued by the JPL, Caltech, under NASA contract 1407.

\citeindexfalse

\end{document}